\documentclass[
 reprint,
superscriptaddress,
 amsmath,amssymb,
 aps,
]{revtex4-2}

\usepackage{graphicx}
\usepackage{dcolumn}
\usepackage{bm}
\usepackage{float}
\usepackage{placeins}
\usepackage{hyperref}
\usepackage{hhline}
\usepackage{multirow}
\usepackage{array}
\usepackage{tabularx}
\usepackage{booktabs}

\begin{document}

\preprint{APS/123-QED}

\title{Sr$_2$IrO$_4$/Sr$_3$Ir$_2$O$_7$ superlattice for a model 2D quantum Heisenberg antiferromagnet}

\author{Hoon~Kim}
\altaffiliation{These authors contributed equally to this work.}
\affiliation{Department of Physics, Pohang University of Science and Technology, Pohang 790-784, Republic of Korea}
\affiliation{Center for Artificial Low Dimensional Electronic Systems, Institute for Basic Science (IBS), 77 Cheongam-Ro, Pohang 37673, South Korea}
\author{Joel~Bertinshaw\footnotemark[1]}
\altaffiliation{These authors contributed equally to this work.}
\affiliation{Max Planck Institute for Solid State Research, Heisenbergstra{\ss}e~1, D-70569 Stuttgart, Germany}
\author{J.~Porras}
\affiliation{Max Planck Institute for Solid State Research, Heisenbergstra{\ss}e~1, D-70569 Stuttgart, Germany}
\author{B.~Keimer}
\affiliation{Max Planck Institute for Solid State Research, Heisenbergstra{\ss}e~1, D-70569 Stuttgart, Germany}
\author{Jungho Kim}
\affiliation{Advanced Photon Source, Argonne National Laboratory 9700 Cass Ave, Lemont, IL 60439, USA}
\author{J.-W.~Kim}
\affiliation{Advanced Photon Source, Argonne National Laboratory 9700 Cass Ave, Lemont, IL 60439, USA}
\author{Jimin~Kim}
\affiliation{Department of Physics, Pohang University of Science and Technology, Pohang 790-784, Republic of Korea}
\affiliation{Center for Artificial Low Dimensional Electronic Systems, Institute for Basic Science (IBS), 77 Cheongam-Ro, Pohang 37673, South Korea}
\author{Jonghwan~Kim}
\affiliation{Center for Artificial Low Dimensional Electronic Systems, Institute for Basic Science (IBS), 77 Cheongam-Ro, Pohang 37673, South Korea}
\affiliation{Department of Materials Science and Engineering, Pohang University of Science and Technology, Pohang 37673, South Korea}
\author{Gahee~Noh}
\author{Gi-Yeop~Kim}
\affiliation{Department of Materials Science and Engineering, Pohang University of Science and Technology, Pohang 37673, South Korea}
\author{Si-Young~Choi}
\affiliation{Department of Materials Science and Engineering, Pohang University of Science and Technology, Pohang 37673, South Korea}
\author{B.~J.~Kim}
\altaffiliation{To whom correspondence should be addressed. \\bjkim6@postech.ac.kr}
\affiliation{Department of Physics, Pohang University of Science and Technology, Pohang 790-784, Republic of Korea}
\affiliation{Center for Artificial Low Dimensional Electronic Systems, Institute for Basic Science (IBS), 77 Cheongam-Ro, Pohang 37673, South Korea}

\date{\today}

\begin{abstract}
Spin-orbit entangled pseudospins hold promise for a wide array of exotic magnetism ranging from a Heisenberg antiferromagnet to a Kitaev spin liquid depending on the lattice and bonding geometry, but many of the host materials suffer from lattice distortions and deviate from idealized models in part due to inherent strong pseudospin-lattice coupling. Here, we report on the synthesis of a magnetic superlattice comprising the single ($n$=1) and the double ($n$=2) layer members of the Ruddlesden-Popper series iridates Sr$_{n+1}$Ir$_n$O$_{3n+1}$  alternating along the $c$-axis, and provide a comprehensive study of its lattice and magnetic structures using scanning transmission electron microscopy, resonant elastic and inelastic x-ray scattering, third harmonic generation measurements and Raman spectroscopy.
The superlattice is free of the structural distortions reported for the parent phases and has a higher point group symmetry, while preserving the magnetic orders and pseudospin dynamics inherited from the parent phases, featuring two magnetic transitions with two symmetry-distinct orders. We infer weaker pseudospin-lattice coupling from the analysis of Raman spectra and attribute it to frustrated magnetic-elastic couplings. Thus, the superlattice expresses a near ideal network of effective spin-one-half moments on a square lattice.  
\end{abstract}

\maketitle

\section{Introduction}
The physics of $S$=1/2 antiferromagnet (AF) on a two-dimensional (2D) square lattice has a long history of research as it is widely believed to hold the key for the mechanism of high temperature superconductivity in copper oxide compounds~\cite{And87,Pat06,Kei15,Pla10,Ore19}. However, it is rarely realized outside of the Cu-based compounds, and as a result its generic features are difficult to isolate from material specifics. Ruddlesden-Popper (RP) series iridates Sr$_{n+1}$Ir$_n$O$_{3n+1}$, have recently emerged as a new material platform to study the same physics with spin-orbit entangled $J_\textrm{eff}$=1/2 pseudospins replacing the $S$=1/2 moments in the cuprates~\cite{Ber19,Kim08,Jac09,Kim09,Wan11}. Indeed, the single layer Sr$_2$IrO$_4$  has reproduced much of the cuprate phenomenology: a pseudogapped metal~\cite{Kim14,Yan15,Cao16}, and a nodal metal with a $d$-wave symmetric gap indicative of possible unconventional superconductivity~\cite{Kim16,Sum17} emerge upon electron doping from the parent phase that is approximately a Heisenberg AF~\cite{Kim12_214,Fuj12}. Further, experiments indicate existence of various symmetry-breaking orders: polarized neutron diffraction~\cite{Jeo17} detects time-reversal symmetry breaking above the N{\'e}el temperature (T$_N$),  magnetic torque~\cite{Mur21} and second harmonic generation~\cite{Zha16,Sey21} measurements indicate loss of C$_4$ rotation symmetry, and resonant x-ray diffraction~\cite{Che18} observes splitting of a magnetic Bragg peak suggesting formation of a nearly commensurate density wave.  

However, iridates are different from the cuprates in several aspects, and to what extent they are relevant to the essential physics is an open important issue. First, the dominant orbital character of the pseudospins leading to strong pseudospin-lattice coupling (PLC)~\cite{Liu19,Porr19,Hu19}, which accounts largely for the spin-wave gap, questions the validity of spin-only models. Second, structural distortions of kinds not found in cuprates~\cite{Fen13, Tor15, Hog16} add complexity to theory models by allowing additional interactions. For example, the staggered tetragonal distortion of IrO$_6$ octahedra in Sr$_2$IrO$_4$, breaking the vertical glide planes and thus lowering the space group from $I$4$_1$/$acd$ to $I$4$_1$/$a$~\cite{Tor15,Fen13}, leads to additional pseudospin exchange interactions, which provide a mechanism for locking of pseudospin canting angles and the octahedral rotation~\cite{Bos13}. In the bilayer compound Sr$_3$Ir$_2$O$_7$, the monoclinic distortion, lowering the space group from orthorhombic $Bbca$ to monoclinic $C$2/$c$~\cite{Hog16} results in bending of otherwise straight Ir-O-Ir $c$-axis bonds. This in turn leads to canting of the AF moments aligned along the $c$-axis, manifesting as small but clearly measurable net in-plane ferromagnetic moments~\cite{Cao02,Nag07}. Such distortions lead to deviation from the ideal cubic-symmetric $J_\textrm{eff}$=1/2 states on rectilinear superexchange pathways, which are assumed in theory models that predict, for example, realization of a Kitaev spin liquid in a honeycomb lattice~\cite{Jac09,Cha10}.

\begin{figure*}[hbt!]
    \centering
    \includegraphics[width=0.95\textwidth ]{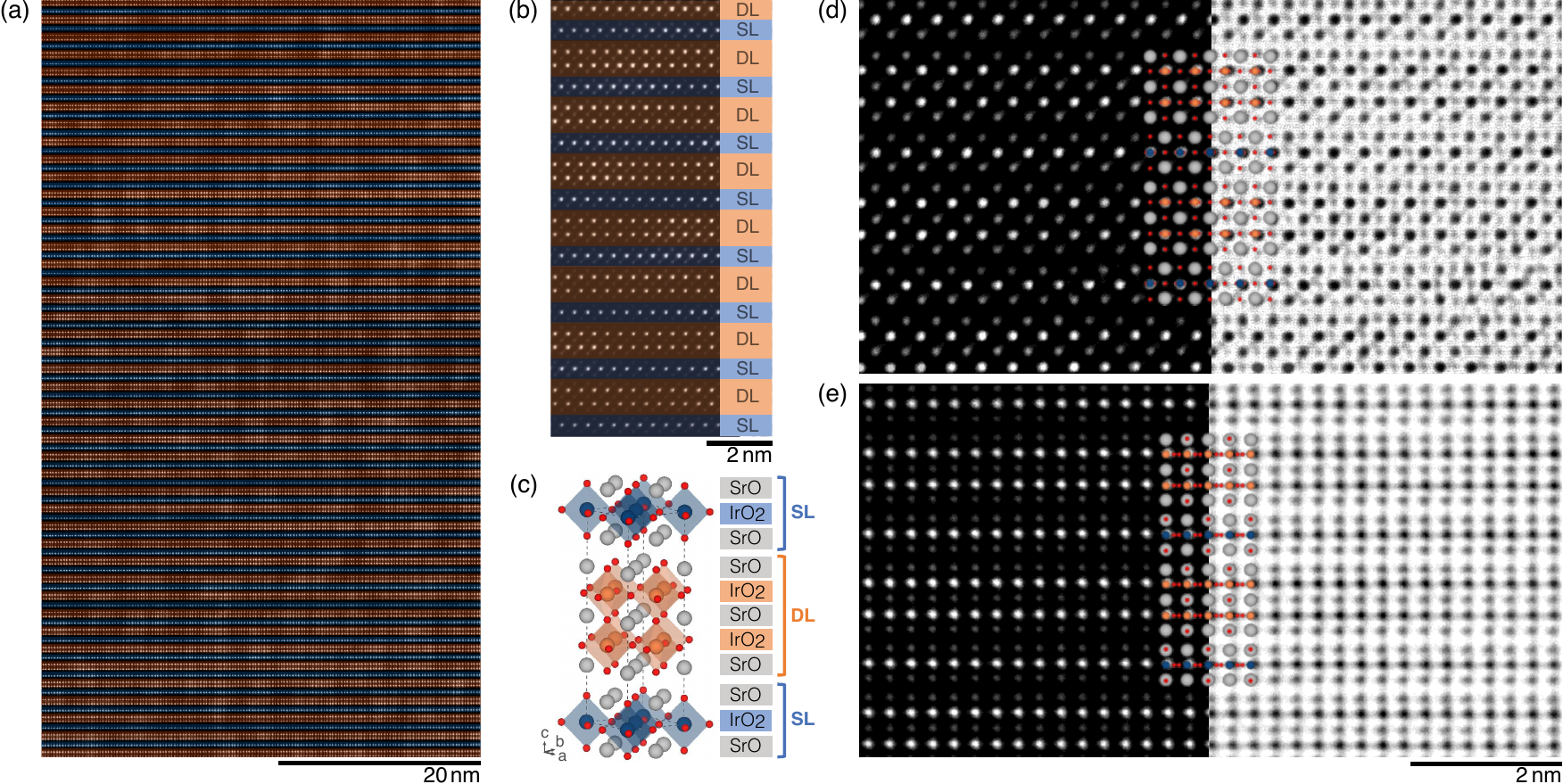}
    \caption{ (color online). 
    Stacking pattern of the superlattice as imaged by STEM.
    (a) Wide field-of-view STEM image along [100] projection. The alternation between single-layers (blue) and double-layers (orange) is well maintained over the entire field of view.
    (b) Magnified HAADF-STEM image with single-layer (blue) and double-layer (orange) indicated.
    (c) A structural model for the superlattice. The single-layer and double-layer are shifted by a half unit cell on the SrO planes. IrO$_6$ octahedra are rotated about the $c$-axis as in the parent compounds. 
    (d) [110]- and (e) [100]-projected HAADF (left)- and ABF (right)-STEM images overlaid with the atom positions (Sr, grey; Ir, blue, orange; O, red) from the model. 
     \label{FIG1}}
\end{figure*}

Here, we report on the synthesis of a Sr$_2$IrO$_4$/Sr$_3$Ir$_2$O$_7$ superlattice, and provide a comprehensive study of its lattice and magnetic structures. The lattice structure is investigated by scanning transmission electron microscopy (STEM), resonant x-ray diffraction (RXD), and rotational anisotropy third harmonic generation measurements (RA-THG), and the magnetic structure by magnetometry, RXD, resonant inelastic x-ray scattering (RIXS), and Raman scattering. The superlattice is free of structural distortions reported for the parent phases while leaving their magnetic structures intact. The superlattice features two magnetic transitions with two different orders: canted $ab$-plane AF and $c$-axis collinear AF inherited from Sr$_2$IrO$_4$ and Sr$_3$Ir$_2$O$_7$, respectively. Their contrasting pseudospin dynamics, of Heisenberg and Ising types, also remain unchanged within our experimental resolutions. However, $ab$-plane magnetic anisotropy of the Heisenberg pseudospins is significantly reduced indicating weaker PLC, possibly due to the Ising pseudospins aligned along the $c$-axis resisting the orthorhombic distortions. Our result shows that two distinct types of quasi-2D magnetism can be compounded in a superlattice to realize a pseudospin system closer to an ideal model.

\section{Lattice Structure}

The superlattice has the nominal chemical composition Sr$_5$Ir$_3$O$_{11}$ and can be regarded as a $n$=1.5 member of the RP series. Although this phase is not quite thermodynamically stable, it forms transiently during a flux growth before either $n$=1 or $n$=2 phase eventually stabilizes depending on the starting composition of Sr$_2$CO$_3$ and IrO$_2$ molten in SrCl$_2$ flux and dwelling temperature. Thermal quenching at high temperature leads to intergrowth of both phases, and a highly ordered superlattice can be found by a careful control of the heating sequence. The resulting ``crystals'' have typically a few microns thickness, thicker than layer-by-layer grown thin films but thinner than typical bulk crystals. As the conventional structure refinement is limited by the small sample volume, we rely on STEM and RXD to verify the superlattice formation.

Figure~\ref{FIG1}(a) shows a wide field-of-view STEM image along [100] projection showing the stacking of single (SL) and double-layer (DL) units alternating over $>$\,40 unit cells. The stacking sequence is indicated in Fig.~\ref{FIG1}(b) and the unit cell is depicted in Fig.~\ref{FIG1}(c), which is modeled from the known structures of Sr$_2$IrO$_4$ (Ref.~\citenum{Craw94}) and Sr$_3$Ir$_2$O$_7$ (Ref. \citenum{Mats04}). Figures~\ref{FIG1}(d) and~\ref{FIG1}(e) show representative high-angle annular dark field (HAADF) and annular bright field (ABF) images along [110] projection and [100] projection, respectively. The images are overlaid with the atomic positions based on the unit cell in Fig.~\ref{FIG1}(c). In the [100] projection, the staggered rotation of IrO$_6$ octahedra about the $c$-axis as in the parent compounds is seen as diffuse rods (see also Figs.~S1 and~S2). Overall, our data is in good agreement with the model.

\begin{figure}[b!]
    \centering
    \includegraphics[width=0.95\columnwidth ]{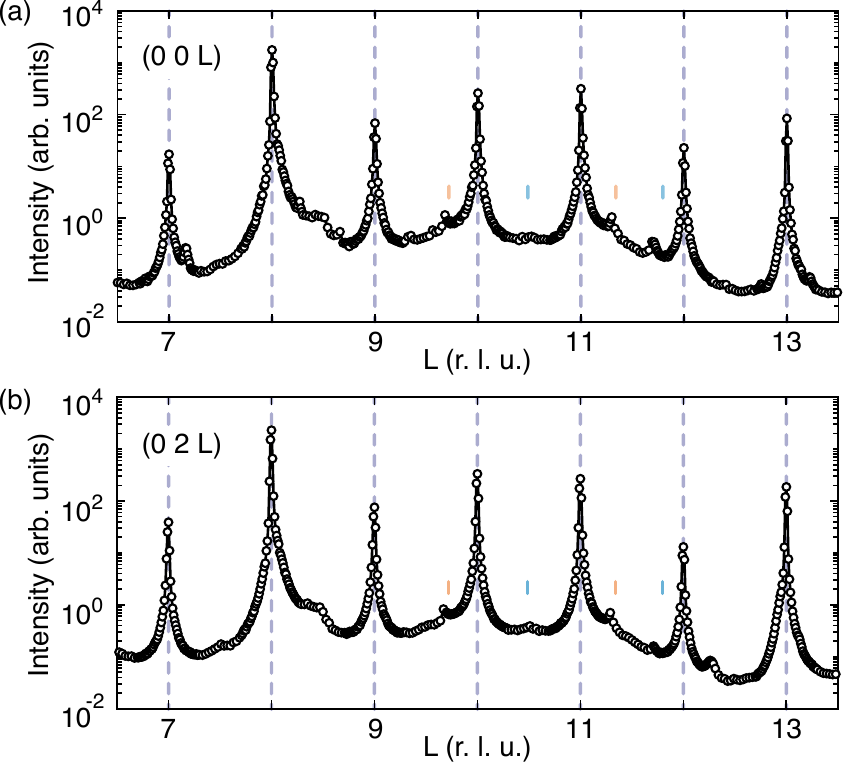}
    \caption{ (color online). RXD on the superlattice at the Ir $L_3$-edge. 
    Sharp (a) (0\,0\,L) and (b) (0\,2\,L) charge reflections are centered at integer-L values (dashed lines), indicating the superlattice is well ordered across the bulk of the sample. Minor impurity peaks that match Sr$_2$IrO$_4$ and Sr$_3$Ir$_2$O$_7$ are marked by cyan and orange sticks, respectively. Miller indices are in the orthorhombic notation; i.e., reciprocal lattice vectors corresponding to the unit cell shown in Fig.~1(c). \label{FIG2}}
\end{figure}

The superlattice formation is confirmed to represent the bulk of the sample via RXD conducted at the Ir $L_3$-edge. Figures~\ref{FIG2}(a) and~\ref{FIG2}(b) plot scans along (0\,0\,L) and (0\,2\,L), respectively, which show reflections centered at integer-L values with $\sim$\,0.01\,\AA\,widths. Impurity peaks of intensities of the order of $\lesssim$\,1\,\% of the main peaks are observed, which match the $c$-axis lattice parameters of either Sr$_2$IrO$_4$ or Sr$_3$Ir$_2$O$_7$. The sharp reflections and negligible impurity peaks indicate that the superlattice structure remains well correlated at a macroscopic level. Whereas the in-plane lattice parameters ($\approx$\,5.502\,\AA) match those of the parent systems, the $c$-axis is unique with a length of 16.93\,\AA. According to the study by Harlow {\it et al.}~\cite{Har95}, however, diffraction patterns from a randomly-stacked intergrowth of $n$=1 and $n$=2 phases can misleadingly appear similar to those of an ordered phase. Such possibility is ruled out in our sample by selectively measuring the periodicity of DL, by exploiting the fact that only DLs are magnetically ordered in the temperature range between T=\,220\,K and 280\,K (to be discussed in section III). 

\begin{figure}[t!]
    \centering
    \includegraphics[width=0.95\columnwidth]{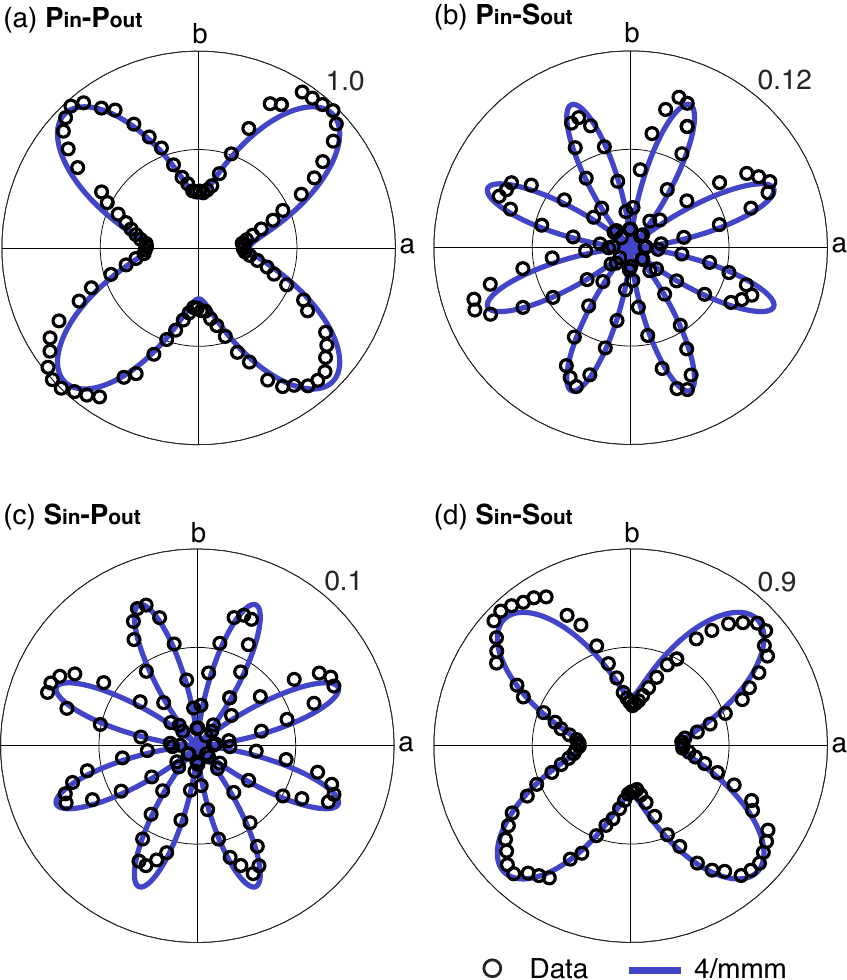}
    \caption{ (color online). RA-THG patterns of the superlattice (open circles) taken under (a) $PP$, (b) $PS$, (c) $SP$, (d) $SS$ geometries. Incident 1200\,nm light was used at room temperature. The third harmonic 400\,nm light was collected as a function of azimuth-angle while the scattering plane rotates about $c$-axis~\cite{Tor14,Har15}. The THG signals are normalized by the $PP$ trace, overlaid with the best fits to bulk electric dipole induced THG tensor of 4/$mmm$ point group (navy lines).
    \label{FIG3}}
\end{figure}

Next, we further refine the structure using RA-THG, a nonlinear optical process highly sensitive to the bulk crystal symmetry through nonlinear susceptibility tensors~\cite{Tor14,Har15,Sip87}. This technique has been used for Sr$_2$IrO$_4$ and Sr$_3$Ir$_2$O$_7$ to detect subtle structure distortions~\cite{Tor15,Hog16,Sey21}. Figure~\ref{FIG3} shows the azimuth-angle dependence of the third harmonic signals as the scattering plane is rotated about the $c$-axis, for the four different polarization configurations of the incident and reflected light, which can be either parallel ($P$) or perpendicular ($S$) to the scattering plane. The patterns are symmetric with respect to mirror reflections $a$$\rightarrow$$-a$ and $b$$\rightarrow$$-b$, and four-fold rotations about the $c$-axis. The combination of both symmetries leads to eight-fold symmetric patterns for the $PS$ and $SP$. To confirm, the patterns are overlaid with the best fit to electric-dipole induced THG tensor for 4/$mmm$ (navy), whose expression is given in Appendix A. We find an excellent agreement and conclude that the superlattice has a higher point group symmetry than Sr$_2$IrO$_4$ (Ref.~\citenum{Tor15}) and Sr$_3$Ir$_2$O$_7$ (Ref.~\citenum{Hog16}), in both of which the patterns manifestly lack the mirror symmetries. 

\section{Magnetic structure}

\begin{figure}[b!]
    \centering
    \includegraphics[width=0.95\columnwidth]{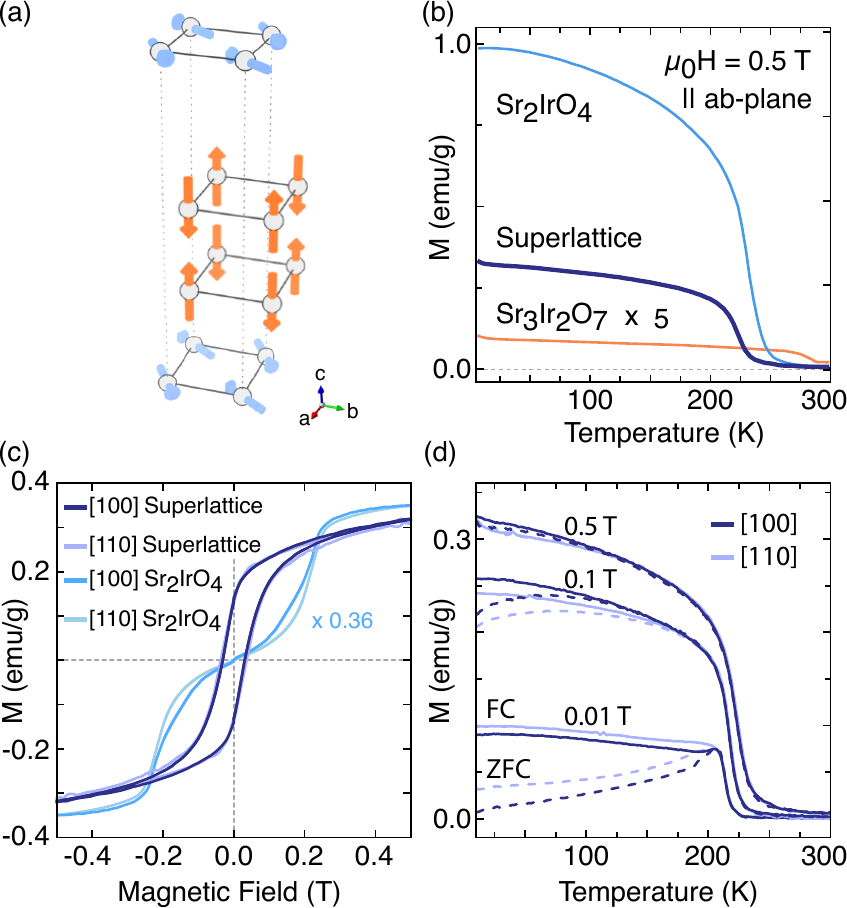}
    \caption{ (color online). Magnetometry of the superlattice. 
    (a) The superlattice magnetic order consistent with our data. 
    (b) Field-cooled M-T curves of the superlattice (navy), Sr$_2$IrO$_4$ (cyan) and Sr$_3$Ir$_2$O$_7$ (orange). 
    (c) M-H hysteresis at 5\,K comparing the superlattice (navy) and the bulk Sr$_2$IrO$_4$ (cyan). For a direct comparison, Sr$_2$IrO$_4$ curves are multiplied by the mass proportion ($\approx$\,0.36) of SL in the superlattice. 
    (d) M-T curves measured with fields applied along [100] and [110].
    \label{FIG4}}
\end{figure}

\begin{figure}[ht!]
    \centering
    \includegraphics[width=0.95\columnwidth]{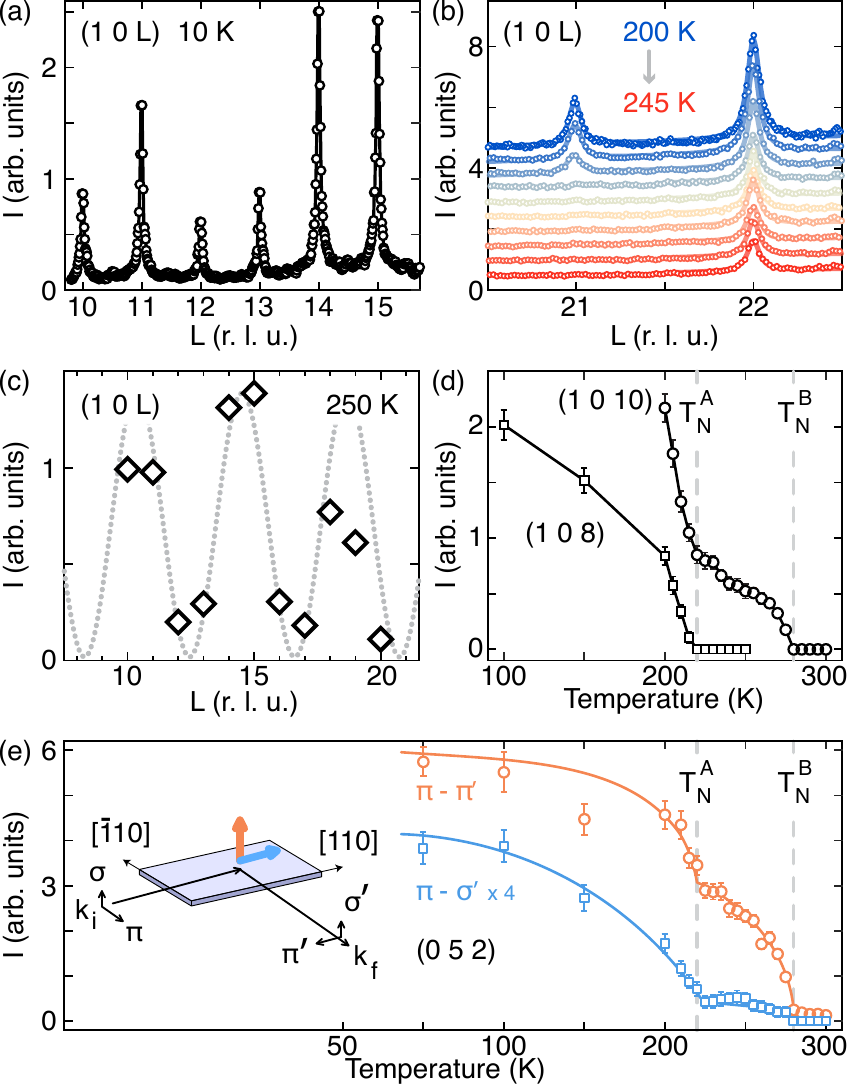}
    \caption{ (color online). Magnetic RXD study on the superlattice.
    (a) Magnetic (1\,0\,L) reflections appear at every integer L values, contributed by signals from both SL and DL.
    (b) (1\,0\,L) scans measured at every 5\,K upon heating from 200\,K to 245\,K. (1\,0\,21) reflection dominated by SL disappears around T\,=\,220\,K.
    (c) At 250\,K, the intensity modulation along (1\,0\,L) coincides with DL structure factor squared (dotted line), indicating that the magnetic intensities are dominated by DL.
    (d) Temperature dependence of (1\,0\,8) and (1\,0\,10) reveal two magnetic transitions at $T_N^A$\,=\,220\,K and $T_N^B$\,=\,280\,K.
    (e) Polarization analysis to separate AF signals from in-plane (blue) and out-of-plane (orange) moments.
    \label{FIG5}}
\end{figure}

Having established the lattice structure of the superlattice, we now turn to its magnetic structure. In short, the magnetic structure remains almost unchanged from the parent systems: SL has $ab$-plane canted AF structure while DL has $c$-axis collinear AF structure, as shown in Fig.~\ref{FIG4}(a).
The net ferromagnetic response to the dc field with the saturation moment close to one-third of that of Sr$_2$IrO$_4$ [Figs.~\ref{FIG4}(b) and~\ref{FIG4}(c)] suggests that the SL (which makes up one-third of the superlattice in terms of the number of Ir ions) has in-plane canted AF structure, while DL has $c$-axis collinear AF structure. We note that the AF ordering in Sr$_3$Ir$_2$O$_7$ is visible as a small jump in the magnetization due to its slight monoclinicity $\beta$\,$\sim$\,90.05\,$^\circ$ (and thereby canting of the moments~\cite{Hog16,Cao02}), but no such anomaly indicative of DL magnetic ordering is seen in our tetragonal superlattice [Fig.~\ref{FIG4}(b)]. Unlike in Sr$_2$IrO$_4$, we observe a ferromagnetic hysteresis loop in the M-H curve shown in Fig.~\ref{FIG4}(c), which implies ferromagnetic staking of SL net moments. Based on our M-H and M-T curves [Figs.~\ref{FIG4}(c) and~\ref{FIG4}(d)] measured for fields along [100] and [110] directions, we are not able to identify the magnetic easy axis in the $ab$-plane, which in Sr$_2$IrO$_4$ is clearly along the $a$-axis~\cite{Porr19,Liu19}. This signifies reduced magnetic anisotropy, which will be discussed in more detail later on. 

\begin{figure*}[t]
    \centering
    \includegraphics[width=0.95\textwidth]{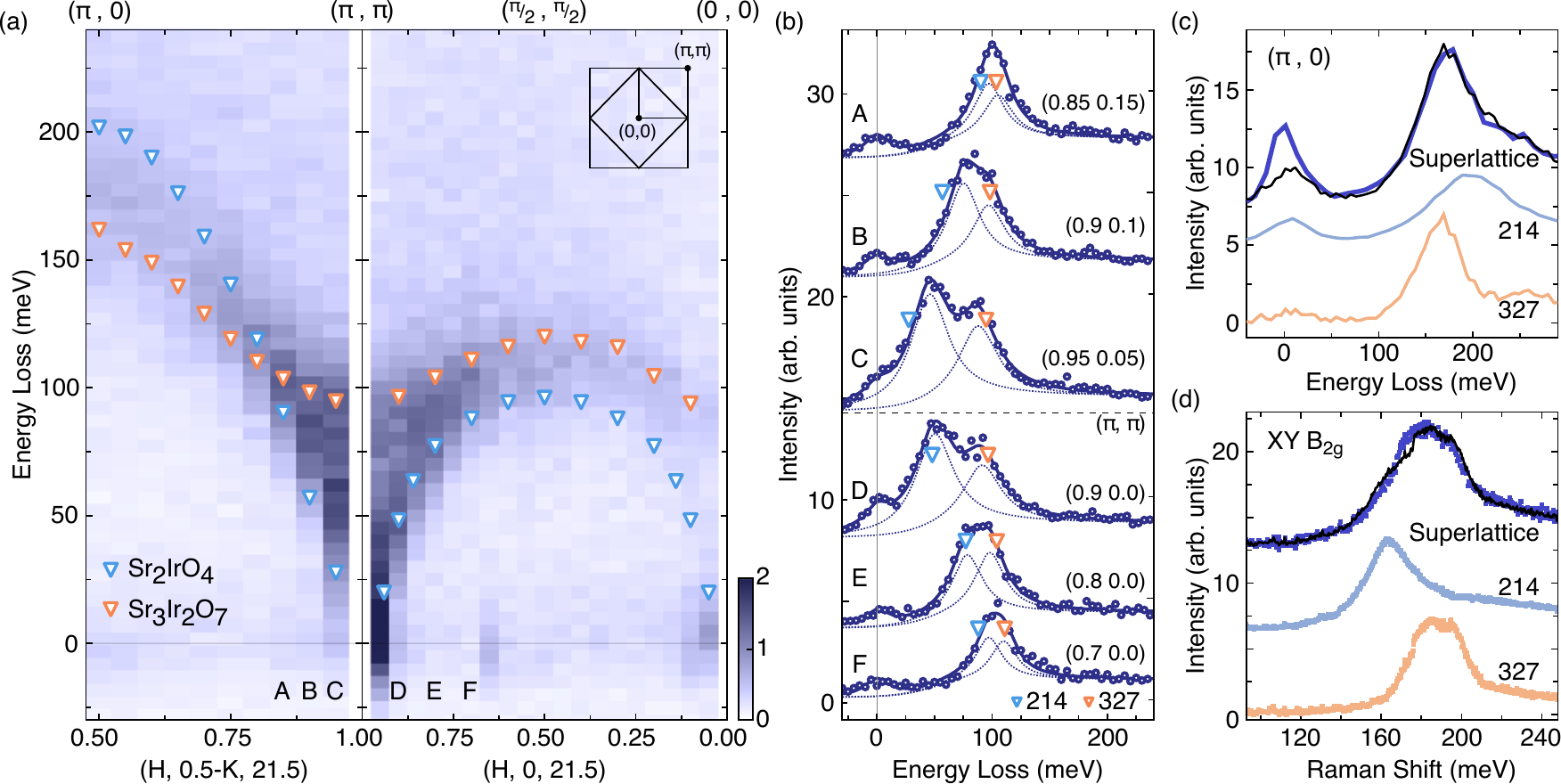}
    \caption{ (color online). Magnetic excitations in the superlattice. 
    (a) RIXS map measured at T\,=10 K along high symmetry directions indicate that the SL and DL modes follow the dispersions of the parent systems (plotted with markers). 
    (b) Spectra at select $\bf{q}$ points. The fitted peaks (solid lines) are compared with those of the parent systems (markers). 
    (c) The superlattice spectrum at the zone corner ($\pi$,0) (navy line) is well reproduced by a linear sum of Sr$_2$IrO$_4$ and Sr$_3$Ir$_2$O$_7$ spectra (black line).
    (d) The two-magnon Raman spectrum (navy dots), measured in the $B_\textrm{2g}$ channel at T\,=\,15 K, is also well approximated by summing the Sr$_2$IrO$_4$ and Sr$_3$Ir$_2$O$_7$ spectral intensities (black line).\label{FIG6}}
\end{figure*}

We confirm the magnetic structure shown in Fig.~\ref{FIG4}(a) using RXD. As in the case of Sr$_2$IrO$_4$ and Sr$_3$Ir$_2$O$_7$, magnetic reflections are found along (1\,0\,L) [Fig.~\ref{FIG5}(a)], but at every integer L, which has contributions from both SL and DL. However, they can be separated by exploiting the DL structure factor. The ratio of Ir-O-Ir bond length along the $c$-axis to $c$ lattice parameter returns oscillation period of $\sim$\,4.15. For example, the DL contribution nearly vanishes for (1\,0\,8) and (1\,0\,21). Indeed, L scans shown in Fig.~\ref{FIG5}(b) shows that (1\,0\,21) peak disappears around T\,=\,220 K as temperature increases while (1\,0\,22) peak is present up to T\,=\,250 K. At this temperature, the intensity modulation well agrees with the DL structure factor squared [Fig.~\ref{FIG5}(c)], implying that the peaks are due to reflections from DL only. This is unambiguous evidence for coherent superlattice formation over the probing depth of x-ray (290\,nm\,$\sim$\,3.1\,$\mu$m as calculated in Ref.~\citenum{Maz21}). The SL transition temperature is measured to be $T_N^A$\,=\,220\,K from the temperature dependence of (1\,0\,8) peak shown in Fig.~\ref{FIG5}(d). At (1\,0\,10), two transitions are seen, the higher temperature one at $T_N^B$\,=\,280\,K being the transition in DL. 

Additional measurements were conducted using polarization analysis in order to separate the $ab$-plane and $c$-axis components of the antiferromagnetic moments. By studying the magnetic (0\,5\,2) reflection in a horizontal scattering geometry [Fig.~\ref{FIG5}(e)], the $\pi$-$\sigma'$ channel mostly detects in-plane moments, whereas $\pi$-$\pi'$ is sensitive to out-of-plane moments. The temperature dependence of the integrated intensities in the two channels, shown in Fig.~\ref{FIG5}(e), reveals that the out-of-plane (in-plane) magnetic signal arises below $T_N^B$\,=\,280\,K ($T_N^A$\,=\,220\,K), thereby located at DL (SL), consistent with the magnetic structure in Fig.~\ref{FIG4}(a).

\section{Pseudospin dynamics}

Having established the static magnetic structure, we now turn to the pseudospin dynamics. Figure~\ref{FIG6}(a) plots the pseudospin excitation spectrum along high symmetry directions. Select $\mathbf{q}$-points are plotted in Fig.~\ref{FIG6}(b). The spectra are fitted with two peaks and their energy positions are compared with the spin-wave dispersions for the parent systems. Overall, the spectra are well described as having two peaks corresponding to spin waves in SL and DL. It is known that Sr$_2$IrO$_4$ is almost gapless at the magnetic zone center reflecting the quasi-continuous symmetry in the $ab$-plane~\cite{Kim12_214}, whereas Sr$_3$Ir$_2$O$_7$ has an exceptionally large gap of $\approx$\,90\,meV (Ref.~\citenum{Kim12_327}). Except in the immediate vicinity of the magnetic zone center, we find a good agreement with the parent systems. In particular, the spectra at the zone corner ($\pi$,\,0) is well reproduced by a linear sum of the spectra of Sr$_2$IrO$_4$ and Sr$_3$Ir$_2$O$_7$ at the same momenta, as shown in Fig.~\ref{FIG6}(c). Further, the two-magnon spectrum measured by Raman scattering [Fig.~\ref{FIG6}(d)], which is dominated by zone-boundary modes, is in perfect agreement with a linear sum of those in Sr$_2$IrO$_4$ and Sr$_3$Ir$_2$O$_7$. Together, these data indicate that the nearest-neighbor spin exchange couplings remain unaltered with respect to the parent systems.

\begin{figure*}[ht!]
    \centering
    \includegraphics[width=0.95\textwidth]{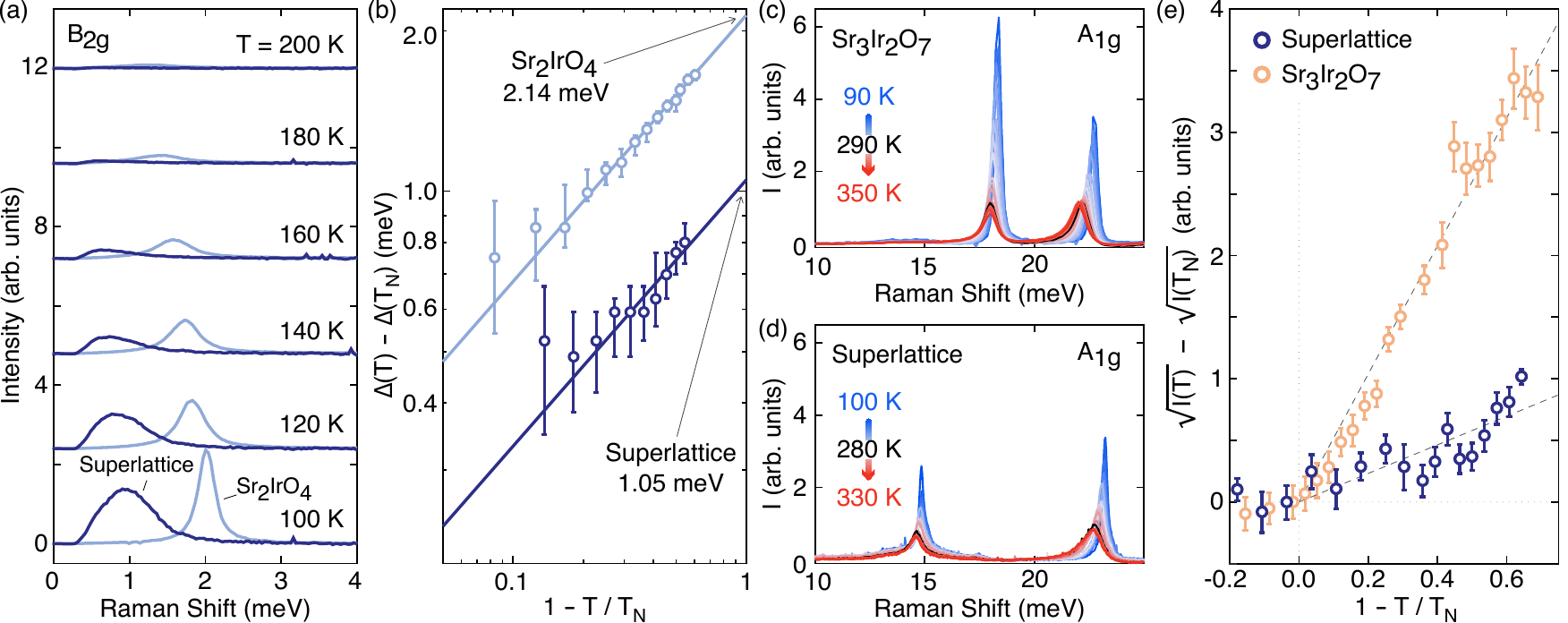}
    \caption{ (color online). Pseudospin-lattice coupling estimated by Raman spectroscopy. (a) The magnetic mode at the zone center is observed in $B_\textrm{2g}$ scattering geometry (superlattice, navy; Sr$_2$IrO$_4$, blue gray). (b) The temperature dependent component of the spin-wave gap $\Delta(T)$ extracted from Raman spectra in (a). The trend of the superlattice and Sr$_2$IrO$_4$ share the same functional form $A\sqrt{1-T/T_N}+B$, where $A$ is the offset in the log-log plot and proportional to the strength of PLC.
    Temperature dependence of $A_\textrm{1g}$ phonons in (c) Sr$_3$Ir$_2$O$_7$ and (d) the superlattice.
    (e) Integrated intensity of the lower energy $A_\textrm{1g}$ phonon. The spin-dependent component scales with the ordered moment squared $M_{AF}^2$, and dashed lines are fits to functional form of $1-(T/T_N)$, whose slopes are proportional to the PLC strength $\Lambda$. All Raman spectra are corrected for laser heating and Bose factors. \label{FIG7}}
\end{figure*}

It is perhaps not surprising that there is no significant change in the pseudospin dynamics across the majority of the Brillouin zone for both SL and DL, considering that these are quasi-2D systems and only the stacking sequence is altered. However, it is rather unusual that the two magnetic subsystems behave almost independently of each other. In particular, magnetic ordering in DL occurs across SL that remains paramagnetic down to $T_N^A$. That no other static order exists in SL above $T_N^A$ is confirmed from its almost identical RIXS spectra measured across $T_N^A$ (Fig.~S4). Our representation analysis (Appendix B) shows that the SL and the DL magnetic orders belong to different irreducible representations, which guarantees that they are not coupled in the linear order (this, however, does not rule out interactions between SL and DL). Further, we note that both $T_N^A$ and $T_N^B$ are reduced only by few degrees from their bulk counterparts, which confirms that interlayer couplings play a minor role in the critical phenomena~\cite{Jac09,Val15,Irk99}.

While the magnetic properties of the superlattice mostly inherit from the parent Sr$_2$IrO$_4$ and Sr$_3$Ir$_2$O$_7$, some differences are also found. Unlike in the case of Sr$_2$IrO$_4$, the in-plane magnetic anistropy of SL is hardly visible in the magnetometry data [Fig.~\ref{FIG4}(c)]. In Sr$_2$IrO$_4$, the magnetic anisotropy is known to arise mostly from PLC~\cite{Liu19,Porr19} stabilizing the magnetic ordering along the $a$- or $b$-axis. The PLC dominates over the nearest-neighbor pseudospin exchange anisotropies favoring alignment along the bond directions (i.e.~along [110]), and also accounts for a dominant part of the in-plane magnon gap~\cite{Porr19}. To compare the strength of PLC, we analyze the temperature dependence of Raman spectra in $B_\textrm{2g}$ symmetry channel, which measures the magnon gap at the zone center [see Fig.~\ref{FIG7}(a)]. It is seen in Fig.~\ref{FIG7}(b) that the magnon gap of both Sr$_2$IrO$_4$ and the superlattice follow the same mean-field-like functional form of $A\sqrt{1-T/T_N}+B$, where the temperature-independent constant $B$ arises from anisotropic magnetic interactions, and $A$ measures the strength of PLC that scales with the size of the ordered moment~\cite{Liu19, Porr19,Hu19}. From the intercept at T\,=\,0 in the log-log plot shown in Fig.~\ref{FIG7}(b), we find that the PLC is drastically reduced by a factor of two. The reduced PLC in SL can be understood in terms of the structural rigidity provided by DL resisting orthorhombic distortions, which would be energetically unfavorable for the $c$-axis collinear AF structure. 

An indication of suppressed PLC can also be found from a phonon mode strongly coupled to the AF order in DL. In Sr$_3$Ir$_2$O$_7$, a recent ultrafast optical spectroscopy study has shown that strong PLC manifests as an abrupt large enhancement across $T_N$ of the amplitude of the coherent oscillation of the impulsive 4.4\,THz ($\approx$\,18 meV) $A_\textrm{1g}$ phonon mode~\cite{Hu19}. The intensity follows the phenomenological relation $\sqrt{I(T)}\propto Z-(\Lambda/4\mu_B^2)M_{AF}^2$, where $\Lambda$ and $Z$ are the PLC strength and temperature-independent spectral weight, respectively~\cite{Suz73}. The strength of PLC in Sr$_3$Ir$_2$O$_7$ is estimated to be two orders of magnitude stronger than in cubic perovskite manganites with strong spin-lattice couplings~\cite{Gra99}. Since the oscillation amplitude is linearly dependent on Raman tensor $\partial\alpha/\partial Q$~\cite{Zei92}, where $\alpha$ is the polarizability and $Q$ is the normal coordinate of the phonon mode, the enhancement should be also visible in Raman spectra. Indeed, we observe a strong enhancement upon cooling of the 18\,meV $A_\textrm{1g}$ mode intensity as shown in Fig.~\ref{FIG7}(c). However, the corresponding mode in the superlattice at $\approx$\,15\,meV shows only a modest increase comparable to that of the 23\,meV mode [Fig.~\ref{FIG7}(e)].  Taken as a whole, the absence of discernible anisotropy in the magnetization data, the reduced magnon gap, and the insensitivty of the $A_\textrm{1g}$ mode to the magnetic order all consistently point to significant reduction of PLC in the superlattice.

\section{Summary}
Recent advances in the control and understanding of 2D materials has recently expanded into the research of magnetism in the extreme quantum 2D limit and in artificial heterostructures. In this work, we have demonstrated a successful growth of a Sr$_2$IrO$_4$/Sr$_3$Ir$_2$O$_7$ magnetic superlattice, whose constituent bulk phases exhibit novel quantum magnetism in the limit of strong spin-orbit coupling.  While intergrowth of different phases in a RP series is frequently found, a natural formation of an ordered superlattice is extremely rare. The superlattice has a lattice symmetry higher than those of the bulk phases and realize an undistorted square lattice geometry. Thus, the superlattice offers a unique opportunity to study the pseudospin physics in an ideal setting. 

The superlattice preserves the bulk magnetic orders and interleaves $ab$-plane canted AF and $c$-axis collinear AF alternately stacked along the $c$-axis. The two mutually orthogonal orders, however, behave independently of each order reflecting weak interlayer couplings expected for quasi-2D systems. In particular, there is a temperature range where only one magnetic subsystem develops an order across the other that remains paramagnetic. Further, the pseudsospin dynamics remains unchanged from the parent systems for the most part of the Brillouin zone. 

Instead, the incompatible nature of the magnetic orders manifests as a strong suppression of the PLC, which is expected to be generally strong for iridates. The reduced PLC in SL is inferred from the smaller zone-center magnon gap, which in Sr$_2$IrO$_4$ is largely accounted for by the PLC through a pseudo-Jahn-Teller effect that results in an orthorhombic distortion as the $ab$-plane magnetic order breaks the tetragonal symmetry of the lattice. This effect, however, is counteracted in the superlattice by DL. The magnetic order in DL is collinear along the straight $c$-axis bond, which in Sr$_3$Ir$_2$O$_7$ is bent due to a slight monoclinicity.

The origin of the distortions in the parent compounds and their absence in the superlattice is a subject of future research. To the best of our knowledge, the breaking of glide symmetries as seen in Sr$_2$IrO$_4$, attributed to staggered tetragonal distortions, is not found in any other transition-metal compounds of the RP type, which suggests that the distortion is not due to an instability of the lattice, but rather its interactions with pseudospins whose magnetic moment size strongly depends on lattice distortions. Similarly, it remains to be understood if PLC plays any role in stablizing the monoclinic structure in Sr$_3$Ir$_2$O$_7$. At any rate, many iridates in bulk form exhibit lattice distortions of some sort and deviate from idealized models, and in this regard the superlattice stands out as a realization of pseudospin one-half physics on an undistorted square lattice. 

The persistent magnetic orders in the superlattice also allows investigation of their evolution upon carrier doping. It is known that Sr$_3$Ir$_2$O$_7$ is on the verge of a metal-insulator transition, and therefore it may well be that DL turns into a Fermi liquid metal while SL remains a correlated fluid, a situation akin to the case of electron-doped Sr$_2$IrO$_4$ via surface alkali metal dosing, where Fermi arcs and a $d$-wave gap are observed, which possibly involve screening effects from the surface metallic layer. Our results presented here provide a solid foundation for these future investigations.     

\section{Methods}

STEM measurements were conducted using JEMARM200F, JEOL at 200\,kV with 5$^{th}$-order probe-corrector (ASCOR, CEOS GmbH, Germany). The specimens were prepared by dual-beam focused ion beam (FIB) with Ga ion beams of 1\,$\sim$\,5\,kV (Nanolab G3 CX, FEI), and further milled by Ar ion beams of 100\,meV (PIPS II, Gatan) to minimize the surface damage. RXD and RIXS measurements were carried out at Beamline 6-ID-B and Beamline 27-ID, respectively, at the Advanced Photon Source, Argonne National Laboratory. For RA-THG measurements, we adapted the scheme reported in~\cite{Har15} by replacing the phase mask by a pair of wedge prisms to accommodate a wider wavelength range of incident light (from 800\,nm to 1200\,nm). The incident 1200\,nm light was provided by an optical parametric amplifier (Orpheus-F, Light Conversion) powered by a femtosecond laser of 100\,kHz repetition rate (Pharos, Light Conversion). Raman spectroscopy is measured with 633\,nm He-Ne laser, which beam of 0.8\,mW is focused on $\sim$2\,$\mu$m.

\section*{Acknowledgement}
This project is supported by IBS-R014-A2 and National Research Foundation (NRF) of Korea through the SRC (no.~2018R1A5A6075964). We acknowledge financial support by the European Research Council under Advanced Grant No.~669550 (Com4Com). The use of the Advanced Photon Source at the Argonne National Laboratory was supported by the U.~S.~DOE under Contract No.~DE-AC02-06CH11357. We acknowledge DESY (Hamburg, Germany), a member of the Helmholtz Association HGF, for the provision of experimental facilities. J.~B. was supported by the Alexander von Humboldt Foundation. G.~Noh, G.-Y.~Kim, and S.-Y.~Choi acknowledge the support of the Global Frontier Hybrid Interface Materials by the NRF (Grant No.~2013M3A6B1078872).
\newline
\section*{APPENDIX A: Third Harmonic Generation Tensor}
We obtain the general form of the bulk electric dipole induced THG tensor for the 4/$mmm$ point group by starting from the most general form of a fourth-rank Cartesian tensor and requiring it to be invariant under all of the symmetry operations contained in the group as:

\begin{widetext}
\begin{equation}
\chi^{ED}_{ijkl} =
\begin{pmatrix} 
\begin{pmatrix}
xxxx & 0 & 0 \\
0 & xxyy & 0 \\
0 & 0 & xxzz
\end{pmatrix} & 
\begin{pmatrix}
0 & xxyy & 0 \\
xxyy & 0 & 0 \\
0 & 0 & 0
\end{pmatrix} & 
\begin{pmatrix}
0 & 0 & xxzz \\
0 & 0 & 0 \\
xxzz & 0 & 
\end{pmatrix} \\

\begin{pmatrix}
0 & xxyy & 0 \\
xxyy & 0 & 0 \\
0 & 0 & 0
\end{pmatrix} & 
\begin{pmatrix}
xxyy & 0 & 0 \\
0 & xxxx & 0 \\
0 & 0 & xxzz
\end{pmatrix} & 
\begin{pmatrix}
0 & 0 & 0 \\
0 & 0 & xxzz \\
0 & xxzz & 0
\end{pmatrix} \\

\begin{pmatrix}
0 & 0 & zxxz \\
0 & 0 & 0 \\
zxxz & 0 & 0
\end{pmatrix} &
\begin{pmatrix}
0 & 0 & 0 \\
0 & 0 & zxxz \\
0 & zxxz & 0
\end{pmatrix} &
\begin{pmatrix}
zxxz & 0 & 0 \\
0 & zxxz & 0 \\
0 & 0 & zzzz
\end{pmatrix}

\end{pmatrix} \;, \tag{A1} \label{eq:A1}
\end{equation}
\end{widetext}
\noindent where $jkl$($i$) stands for incident(scattered) light polarizations. In our experiment, the light is incident on the sample surface normal to the $c$-axis with the incidence angle $\theta$ and azimuth angle $\psi$, which is defined to be zero when the scattering plane contains the $a$ and $c$ axes). Then, the polarization vectors are given by

\begin{align}
    \vec{\epsilon}_{s}\,&=\,(\sin\psi,\, -\cos\psi,\, 0)\;, \notag \\
    \vec{\epsilon}_{in,p}\,&=\,(-\cos\theta\cos\psi,\, -\cos\theta\sin\psi,\, \sin\theta)\;, \tag{A2} \\
    \vec{\epsilon}_{out,p}\,&=\,(\cos\theta\cos\psi,\, \cos\theta\sin\psi,\,\sin\theta)\;, \notag
\end{align}

Multiplying the tensor with the polarization vectors, the expressions for the THG intensity simplify as 

\begin{align}
    I^{SS}(3\omega)\,\sim\,&|A + B\cos(4\psi) |^2\;, \notag \\
    I^{PS}(3\omega)\,\sim\,&|B \sin(4\psi) |^2\;, \tag{A3} \\
    I^{SP}(3\omega)\,\sim\,&|B \sin(4\psi) |^2\;, \notag \\
    I^{PP}(3\omega)\,\sim\,&|A' + B\cos(4\psi)|^2\;, \notag
\end{align}

\noindent where $A$, $A'$ and $B$ are adjustable parameters. The formulae are consistent with the 4/$mmm$ point group symmetry, invariant under both mirror reflections ($\psi$ $\rightarrow$ $\pi-\psi$ and $\psi$ $\rightarrow$ $-\psi$) and C$_4$ rotation about the $c$-axis ($\psi$ $\rightarrow$ $\pi/2+\psi$). The full expressions and those for lower symmetry point groups are given in the supplemental materials.

\begin{table*}[ht!]
\centering
\begin{tabular}{>{\centering\arraybackslash}m{1cm}>{\centering\arraybackslash}m{2.5cm}>{\centering\arraybackslash}m{2.5cm}>{\centering\arraybackslash}m{0.5cm}>{\centering\arraybackslash}m{2.5cm}>{\centering\arraybackslash}m{2.5cm}>{\centering\arraybackslash}m{2.5cm}>{\centering\arraybackslash}m{2.5cm} } 
\hhline{========} \\
\multirow{ 2}{*}{Irrep.} & 
\multicolumn{2}{c}{\bf{Single layer}} & & 
\multicolumn {4}{c}{\bf{Double layer}} \\ [\medskipamount]
 & a1 & a2 & & b1 & b2 & b3 & b4\\ [\medskipamount] \cline{2-3}\cline{5-8} \\

\multirow{2}{*}{$\Gamma_1$} & 
(m$_1$,\;m$_2$,\;0) & (-m$_1$,\;m$_2$,\;0) & &
(m$_3$,\;m$_4$,\;0) & (m$_5$,\;m$_6$,\;0) & (-m$_3$,\;m$_4$,\;0) & (-m$_5$,\;m$_6$,\;0) \\
& (-m$_2$,\; m$_1$,\;0) & (-m$_2$,\;-m$_1$,\;0) & &
(-m$_6$,\;-m$_5$,\;0) & (-m$_4$,\;-m$_3$,\;0) & (-m$_6$,\;m$_5$,\;0) & (-m$_4$,\;m$_3$,\;0) \\ [\bigskipamount]
$\Gamma_2$ &
 & & & (0,\;0,\;m$_1$) & (0,\;0,\;-m$_1$) & (0,\;0,\;m$_1$) & (0,\;0,\;-m$_1$) \\ [\bigskipamount]
$\Gamma_3$ &
 & & & (0,\;0,\;m$_1$) & (0,\;0,\;m$_1$) & (0,\;0,\;-m$_1$) & (0,\;0,\;-m$_1$) \\ [\bigskipamount]
$\Gamma_4$ &
(0,\;0,\;m$_1$) & (0,\;0,\;m$_1$) & & (0,\;0,\;m$_2$) & (0,\;0,\;m$_2$) & (0,\;0,\;m$_2$) & (0,\;0,\;m$_2$) \\ [\bigskipamount]
$\Gamma_5$ &
(0,\;0,\;m$_1$) & (0,\;0,\;-m$_1$) & & (0,\;0,\;-m$_2$) & (0,\;0,\;m$_2$) & (0,\;0,\;m$_2$) & (0,\;0,\;-m$_2$)
 \\ [\medskipamount]
\hhline{========}
\end{tabular}
\caption{Irreducible representations and basis vectors for the magnetic structures allowed in the superlattice. m$_i$ (i$=1,\cdots,6$) are independent parameters for the magnetic moments. Iridium ions are located at a1:(0,0,0), a2:(1/2,1/2,0), b1:(0,1/2,$\delta$), b2:(0,1/2,1-$\delta$), b3:(1/2,0,$\delta$), b4:(1/2,0,1-$\delta$) in the unit cell Fig.~\ref{FIG1}(c), where $\delta$\,=\,0.3791. The magnetic structure in Fig.~\ref{FIG4}(a) comprises canted $ab$-plane AF of $\Gamma_1$ and $c$-axis collinear AF of $\Gamma_5$.
\label{TAB1}}
\end{table*}

\section*{APPENDIX B: Representation Analysis of the magnetic structure}

We present a representation analysis based on the lattice structure shown in Fig.~\ref{FIG1}(c), assuming $\mathbf{q}$\,=\,0 propagation vector based on the fact that all observed magnetic reflections have integer Miller indices. Its space group is $P\bar{4}b$2, which lacks inversion symmetry and thus its point group ($\bar{4}m2$) is of lower symmetry than 4/$mmm$. RA-THG is, however, insensitive to the inversion symmetry and has the same tensor structure for the two point groups. The inversion symmetry is broken by the way in which octahdera are rotated in DL in the current model. An inversion symmetric structure model can be made by doubling the unit cell in such a way that the octahedral rotation is opposite on two neighboring DLs. Thus, the determination of the presence of inversion symmetry requires full structure refinement including the octahedral rotation on each layers, which is beyond the scope of this work. However, our result that the magnetic orders in SL and DL belong to different irreducible representations (IR) is not affected by these details. 

In the superlattice, iridium ions in SL and DL are not symmetry related and thus their magnetic structures are analyzed separately. The result of the analysis is shown in Table~\ref{TAB1}. In both SL and DL, the $ab$-plane and $c$-axis components belong to different irreducible representations. This can easily be seen from the transformation property under the two-fold rotation about $c$-axis; the $ab$-plane moments are flipped, whereas the $c$-axis moments remain invariant. As long as the $ab$-plane and $c$-axis components are not mixed by any of the symmetry operations contained in the space group, their different characters for the two-fold rotation guarantee that they belong to different IRs. A more detailed version of the analysis is presented in the supplemental material.

\providecommand{\noopsort}[1]{}\providecommand{\singleletter}[1]{#1}%

\end{document}